\documentclass[twocolumn,showpacs,preprintnumbers,amsmath,aps]{revtex4}
\usepackage{epsfig}

\begin{document}

\title{Rigorous solution of the spin-1 quantum
Ising model with single-ion anisotropy}

\author{ Zhihua Yang$^{1}$, Liping Yang$^2$, Jianhui Dai$^{1,3}$,
and Tao Xiang$^{4,2}$}

\address{$^1$Zhejiang Institute of Modern Physics, Zhejiang University,
Hangzhou 310027,  China\\
$^2$Institute of Theoretical Physics, Chinese Academy of
Science, P.O. Box 2735, Beijing 100080, China\\
$^3$Department of Physics and Astronomy, Rice University, Houston
TX-77005, USA\\
$^4$Institute of Physics, Chinese Academy of Sciences, P.O. Box 603,
Beijing 100080, China }

\date{\today}

\begin{abstract}

We solve the spin-1 quantum Ising model with single-ion anisotropy
by mapping it onto a series of segmented spin-1/2 transverse Ising
chains, separated by the $S^z =0$ states called holes. A recursion
formula is derived for the partition function to simplify the
summation of hole configurations. This allows the thermodynamic
quantities of this model to be rigorously determined in the
thermodynamic limit. The low temperature behavior is governed by the
interplay between the hole excitations and the fermionic excitations
within each spin-1/2 Ising segment. The quantum critical
fluctuations around the Ising critical point of the transverse Ising
model are strongly suppressed by the hole excitations.
\end{abstract}

\maketitle

The quantum lattice models for which both the ground state and the
finite temperature thermodynamics can be exactly solved are rare.
However, these models play an important role in the study of quantum
criticality. A typical example is the uniform $S=1/2$ transverse
Ising model(TIM), which sets a paradigm in elucidating the nature of
both quantum and thermodynamic phase
transitions\cite{Lieb61,Pfeuty,Sondhi,Sachdev,Kopp}. Experimentally,
this kind of lattice models can be realized in certain electronic
materials\cite{Bitko,Chakrabarti,Richter} as well as in the optical
lattice of cold atoms or polar molecules\cite{Duan,Zoller}. However,
in many realistic situations, the local moments are larger than
$1/2$ and exposed to the interaction of single-ion anisotropy
generated by the crystal fields\cite{Abragam}. In particular, in an
integer spin system, a local spin can be in a neutral polarized
states with $S^z=0$. It is unclear how these extra degrees of
freedom of spins can change the nature of quantum criticality
revealed by the S=1/2 TIM. To address unambiguously this problem, it
is desired to find an extended but exactly soluble TIM model with
higher spins.

In this Letter, we study an Ising model of S=1 with the single-ion
anisotropy, defined by the following Hamiltonian
\begin{equation}
H=-\sum_{j=1}^L \left[ J S_{j}^z S_{j+1}^z + 2 D_x (S_{j}^{x})^2 +
D_z (S_{j}^{z})^2 \right], \label{eq:ham}
\end{equation}
where $L$ is the lattice length. In the classical limit, i.e
$D_x=0$, it reduces to the Blume-Capel model\cite{Blume66,Capel66}.
The ground state of this model has been studied by a number of
authors\cite{Eddeqaqi,Oitmaa}. Oitmaa and Brasch first pointed out
that in the ground state this model is equivalent to the $S=1/2$
TIM\cite{Oitmaa} and can therefore be exactly solved. Here we want
to show that the thermodynamic quantities of this model can be also
rigourously calculated. To our knowledge, it is the first quantum
$S=1$ spin model whose thermodynamic properties can be rigourously
studied without invoking the Bethe Ansatz.

In Eq.~(\ref{eq:ham}), $D_x$ and $D_z$ are the coupling constants of
the single-ion anisotropy along the $x$- and $z$-axes, respectively.
A single-ion anisotropy along the $y$-axis can be added to this
Hamiltonian. However, this term is not independent since
$\mathbf{S}^2=2$. It can be absorbed into the $D_x$ and $D_z$ terms.
Eq.~(\ref{eq:ham}) can be also extended to include the spin-1/2
magnetic impurities as well as longitude magnetic field. In the
discussion below, free boundary conditions are assumed. It is
straightforward to extend the results to the system with periodic
boundary conditions.

At each site of the lattice, $S_j^z$ can take three values, $S_j^z=
0, \pm 1$. Effectively, one can regard $S_j^z =\pm 1$ as the two
polarized spin states of a S=1/2 spin operator and $S_j^z=0$ as a
hole. A remarkable property of the Hamiltonian is that at each site
$(S_j^z)^2$ commutes with $H$. This means that the hole states
($S_j^z= 0$) are decoupled from the spin polarized states ($S_j^z =
\pm 1$). Thus the total number of holes is a good quantum number and
can be used to classify the eigenstates of $H$.

The holes in this system act like non-magnetic impurities. They
will separate the system into many independent segments of
interacting S=1/2 spins. In a system of $p$ holes, there are at
most $p+1$ segments of S=1/2 spins. If these holes are located at
$\{ x_1, \cdots x_p\}$ with $1\leq x_1<\cdots <x_p\leq L$, it is
straightforward to show that Eq. (\ref{eq:ham}) is exactly
equivalent to the following Hamiltonian up to a dynamic irrelevant
constant (setting $x_{0} = 0$ and $x_{p+1} = L+1$)
\begin{equation}
H(\{x_i,p\}) =  \sum_{n=1}^{p+1}h(l_n) + p(D_z - D_x) ,
\label{eq:hp}
\end{equation}
where
\begin{equation}
h(l_n)  = -\sum_{j=x_{n-1}+1}^{x_{n}-2} J  \sigma_{j}^z
\sigma_{j+1}^z - \sum_{j=x_{n-1}+1}^{x_{n} - 1} D_x \sigma_{j}^x ,
\label{eq:hsegment}
\end{equation}
$\sigma_\mu\, (\mu = x, z)$ are the Pauli matrices and
$l_n=x_{n}-x_{n-1}-1$ is the segment length.

The above discussion indicates that in order to diagonalize the
model, one needs only to diagonalize the Hamiltonian for each
individual segment defined by Eq. (\ref{eq:hsegment}). The
eigenfunction of $H$ is a product of all the eigenfunctions of the
segment Hamiltonians $h(l_n)$. Correspondingly, the eigenvalue of
$H$ is simply given by the sum of the eigenvalues of $h(l_n)$.
$D_z-D_x$ plays the role of chemical potential of holes. Thus by
adjusting the value of $D_z$, one can control the number of holes in
the ground state.

$h(l)$ defined by Eq. (\ref{eq:hsegment}) is the Hamiltonian of the
S=1/2 Ising model in a transverse field. It can be reduced to a
model of non-interacting fermions by the Jordan-Wigner
transformation. By further diagonalizing this fermionic model, the
excitation spectrum can be obtained. The energy dispersion of the
fermionic excitation is given by
\begin{equation}
\varepsilon(l )=\pm|D_x|\sqrt{1+\lambda^2+2\lambda\cos k} ,
\label{eq:dispersion}
\end{equation}
where $\lambda= J/D_x$ and $k$ is determined by the secular
equation:
\begin{eqnarray}
\frac{\sin(l +1)k}{\sin l k} = -\lambda. \label{eq:secular}
\end{eqnarray}
This equation is reflection symmetric: if $k$ is a solution, then
$-k$ is also a solution. For $|\lambda | < 1$, $k$ has $2l$ real
roots within the interval $(-\pi,\pi ]$. However, for $|\lambda |
> 1$, $k$ has $2l-2$ real roots within $(-\pi,\pi ]$ and two
opposite complex roots. The eigenstates corresponding to these two
complex solutions of $k$ are two localized states, each trapped at
one end of the chain.

\begin{figure}[ht]
\centerline{\includegraphics [bb=0 0 303 240 ,width=0.9\columnwidth]
{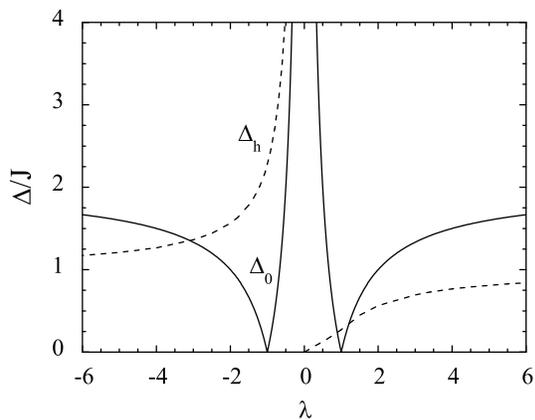}} \caption[] {Minimal energy for exciting a fermion
quasiparticle $\Delta_0$ or a hole $\Delta_h$ when $D_z = 0$. }
\label{fig:gap}
\end{figure}

The Hamiltonian (\ref{eq:ham}) contains two kinds of excitations.
One is the fermionic excitation within each segment and the other is
the hole excitations. The fermion excitation is gapped except at the
critical point $| \lambda | = 1$. The minimal excitation gap is
given by
\begin{equation}
\Delta_0 = 2 |J| \left| \frac{1}{|\lambda |} -1 \right|.
\label{eq:gap0}
\end{equation}
The minimal hole excitation gap is determined by the minimal
energy for creating a hole at one end of the spin chain and given
by
\begin{equation}\label{eq:gaph}
\Delta_h = E_0(L-1) - E_0(L) + D_z - D_x ,
\end{equation}
where $E_0(l)$ is the ground state energy of $h(l)$. Eq.
(\ref{eq:gaph}) holds when $\Delta_h > 0$. In the case $\Delta_h <
0$, holes will appear in the ground state. These holes will condense
and break the system into many S=1/2 spin segments. In this case,
the hole excitation becomes gapless.

Fig. (\ref{fig:gap}) shows the $\lambda$ dependence of the fermion
and hole excitation gaps $\Delta_0$ and $\Delta_h$ for $D_z =0$. The
spectrum of $h(L)$ is unchanged when $\lambda$ changes to
$-\lambda$. Thus $\Delta_0$ is symmetric under the refection of
$\lambda$. However, the hole excitation gap is non-symmetric when
$\lambda$ changes to $-\lambda$. When $\lambda >0$, $\Delta_h$ is
below $\Delta_0$ except in a narrow region around the critical point
$\lambda = 1$. At the critical point $\lambda = 1$, $\Delta_h =
0.273J$. In the case $\Delta_h < \Delta_0$, the low energy
excitations are dominated by the hole excitations.

When $D_z =0$, there is no hole in the ground state. In this case,
the ground state can be either in a spin ordered phase at which
all spins are either ferromagnetically or antiferromagnetically
polarized along the z-axis depending on the sign of $J$, or in a
quantum disordered phase without any long-range spin order. The
transition between these two phases at $|\lambda| = 1$ is of the
Ising criticality.

Now let us consider the thermodynamic properties of the model. At
first glance, it seems to be extremely difficult to calculate
rigourously thermodynamic quantities of this system since the
holes can take exponentially many configurations even though
$h(l)$ can be analytically solved. However, for the system studied
here, we find that the partition function can be expressed by the
following recursion formula
\begin{equation}\label{eq:partition}
Z = \textrm{Tr}  \exp \left( -\beta H \right) = \sum_{p=0}^L
\alpha^p Z^{(p)}(L-p)
\end{equation}
where $\alpha = \exp\left[ \beta (D_x - D_z ) \right] $ and
\begin{eqnarray}
Z^{(p)}(l) & = & \sum_{l_1 + \cdots l_{p+1} = l } z(l_1) z(l_2)
\cdots z(l_{p+1})\nonumber \\
& = & \sum_{l_{1}=0}^{l} z(l_{1}) Z^{(p-1)} (l -l_{1}).
\label{eq:iter}
\end{eqnarray}
Here, we define $Z^{(0)}(l) = z(l) = \textrm{Tr} \exp\left[- \beta h
(l) \right]$,  $z(0) = 1$, and $z(1)=2 \cosh(\beta D_x)$. Thus the
partition function can be evaluated recursively with Eq.
(\ref{eq:iter}), starting from a no-hole system. Given $z(l)$, the
computing time needed just increases quadratically instead of
exponentially with the lattice size. Therefore, this has greatly
simplified the calculation. It allows us to access readily the
thermodynamic limit by evaluating exactly the thermodynamic
quantities in a sufficiently large lattice, for example $L=10^4$, at
which the finite size effect can be ignored. 

\begin{figure}[ht]
\centerline{\includegraphics [bb=5 11 344 428
,width=0.9\columnwidth] {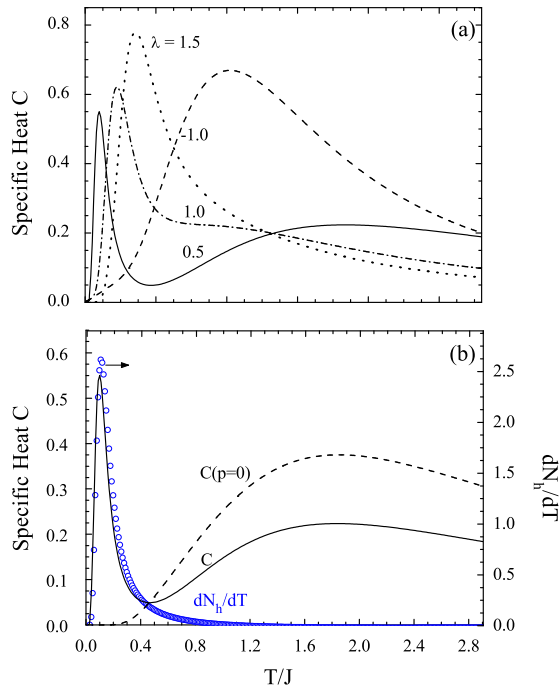}} \caption[] {Temperature
dependence of (a) the specific heat $C$ for several different
$\lambda$ and (b) the specific heat $C$ and the temperature
derivative of the hole excitation number $dN_h/dT$ for the $S=1$
Ising spin model (\ref{eq:ham}) with $\lambda = 0.5$, $J = 1$ and
$D_z = 0$. The specific heat for the same model but without hole
excitations, $C(p=0)$), is also shown for comparison. }
\label{fig:heat}
\end{figure}

From the partition function and its temperature derivatives, one can
evaluate the free energy and all other thermodynamic quantities.
However, one can also calculate directly the internal energy, the
correlation functions, and other measurable variables from the
corresponding segment quantities using the recursion formula of the
partition function. For example, the internal energy is given by
\begin{eqnarray}
U & = &  \sum_{p=0}^L \alpha^p (p+1) \sum_{l = 0}^{L-p}\frac{ u(l)
z(l) Z^{(p-1)}(L-p-l)}{Z(L)} \nonumber \\
&& + \left(D_z - D_x \right) N_h , \label{eq:expect1}
\end{eqnarray}
where, $Z^{(-1)}(l)\equiv \delta_{l,0}$, $u(l) = z^{-1}(l)
\textrm{Tr} h (l) \exp [-\beta h(l)]$ is the internal energy of a
segment, and $N_h$ is the thermal average of the hole number
defined by
\begin{equation}
N_h =\frac{1}{Z(L)} \sum_{p=0}^L p \alpha^p Z^{(p)}(L-p).
\end{equation}
The specific heat can then be determined from the first order
derivative of the internal energy.

Fig.~(\ref{fig:heat}) shows the temperature dependence of the
specific heat for several different $\lambda$ and the temperature
derivation of the hole excitation number. When $T \ll \min
(\Delta_0, \Delta_h)$, the specific heat drops to zero exponentially
with decreasing temperature, except at the quantum critical points.
Above this exponential temperature dependent regime, two low
temperature peaks appear in the specific heat curves when $0 <
\lambda < 1.5$. These two peaks result from the hole and fermionic
excitation and appear roughly at $T \sim \Delta_h$ and $T \sim
\Delta_0$, respectively. This in fact can be more clearly seen from
Fig.~(\ref{fig:heat}-b) where the temperature dependence of the
specific heat for $\lambda = 0.5$ is shown and compared with the
corresponding curve without any hole excitation $C (p = 0)$ and with
the temperature derivative of the thermal average number of holes
$\textrm{d} N_h / \textrm{d} T$. For $\lambda = 0.5$, $\Delta_h <
\Delta_0$, the low-lying excitations are dominated by the hole
excitations. Thus the low temperature peak arises from the hole
excitations.

\begin{figure}[ht]
\centerline{\includegraphics [bb=0 0 321 240, width=0.9\columnwidth]
{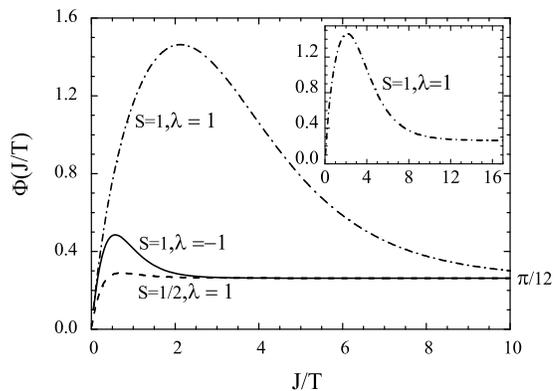}} \caption[] {The temperature dependence of $\Phi (T)$.
The inset shows $\Phi (T)$ versus $J/T$ for $\lambda =1$ and $S=1$
in a wider temperature range.} \label{fig:scaled}
\end{figure}

When $D_z=0$, there are no holes in the ground state. Thus the
quantum critical behavior of the model is not affected by the hole
excitations at zero temperature. However, at finite temperature, the
hole excitations can enhance the thermal fluctuation and suppress
strongly the onset temperature $T^*$ below which the quantum
criticality is observed around the quantum critical point. Recently,
Kopp and Chakravarty  calculated the scaled free energy coefficient
\cite{Kopp} for the $S=1/2$ TIM
\begin{equation}\label{eq:free}
\Phi (T) =\frac{2|J|\sqrt{|\lambda|} [F(0) -F(T)]}{T^2},
\end{equation}
where $F(T)$ is the free energy, and the factor
$2|J|\sqrt{|\lambda|}$ is the velocity of the elementary
excitations ($\hbar=1$) at low energy. At the critical point,
$\lambda =1$, they found that $\Phi (T)$ already falls into the
quantum critical scaling regime with $\Phi (T) \approx \Phi (0) =
\pi / 12$ at $T^* \lesssim J/2$. This suggests that the quantum
criticality in the $S=1/2$ TIM can persist up to a surprisingly
high temperature. However, in the presence of holes, we find that
the persistence of the quantum critical behavior of the system is
modified. Fig.~(\ref{fig:scaled}) shows $\Phi$ as a function of
$J/T$ for the $S=1$ TIM at the two critical points ($\lambda =\pm
1$) and compared with the corresponding results for the $S=1/2$
TIM. In the case $\lambda = 1$, $\Phi$ shows a broad peak around
$T \sim J/2$ and the onset temperature $T^*$ of quantum
criticality is suppressed by nearly one order of magnitude due to
the hole excitations. However, at $\lambda=-1$, the suppression is
relatively weak since the hole excitation gap is much larger than
$J/2$.

When the hole excitation has a finite energy gap, the ground state
is magnetically disordered when $|\lambda | < 1$ but ordered (either
ferromagnetic or antiferromagnetic, depending on the sign of $J$)
when $|\lambda |> 1$. However, for sufficiently large but negative
$D_z$, $\Delta_h$ defined by Eq. (\ref{eq:gaph}) becomes negative.
In this case, the static holes will condense in low temperatures.
Fig.~(4-a) shows the ground state phase diagram. Across the hole
condensed phase boundary, both the magnetization $m$ and the hole
excitation number $N_h$, as shown in Fig. (4-b,c), change
discontinuously at zero temperature. This is a typical
characteristic of the first order phase transition.

\begin{figure}[t]\label{fig:phase}
\centerline{\includegraphics [bb=5 11 221 257,width=0.9\columnwidth]
{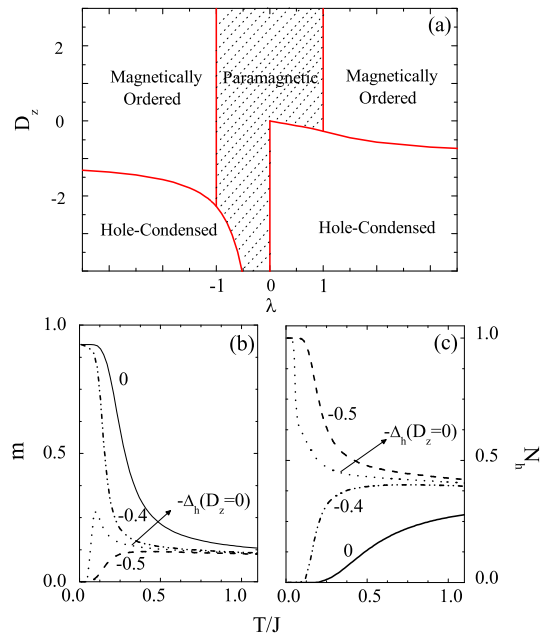}} \caption[] {(a) The phase diagram of the ground state.
(b) and (c) show the temperature dependence of the magnetization
$m=\sqrt{\sum_{ij}\langle S_i^zS_j^z\rangle}/L$ and the
corresponding hole excitation number $N_h$ for several $D_z$ with
$J=1$, $\lambda=1.5$, and $\Delta_h(D_z=0)=0.447974J$.}
\end{figure}

In summary, we have developed a recursive method to solve the
thermodynamics of the $S=1$ quantum Ising chain with single-ion
anisotropy. This allows us to evaluate rigorously all thermodynamic
quantities in the thermodynamic limit. The hole excitations affect
strongly the low temperature behaviors of the system. They enhance
the thermodynamic fluctuations and reduce strongly the
characteristic temperature of quantum criticality. The formula
derived in this work hold not just for the model studied here. With
proper extension, they can also be applied to study thermodynamic
properties of quasi-one dimensional antiferromagnets with
nonmagnetic impurities, such as
Sr$_2$(Cu$_{1-x}$Pd$_x$)O$_3$\cite{Kojima,Sirker},
Cu$_{1-x}$Zn$_x$GeO$_3$\cite{Hase}, as well as other physical
systems whose Hamiltonian can be written as a sum of independent
spin segments, separated by non-magnetic impurities.

This work was supported by the NSFC, the national program for basic
research of China, PCSIRT and SRFDP (J20050335118)of the Chinese
Ministry of Education.

\end{document}